# Requirements for a career in information security: A comprehensive review

*Mike Nkongolo[1], Nita Mennega [1], Izaan van Zyl[1]
[1]University of Pretoria, Department of Informatics, South-Africa

[1]**mike.wankongolo@up.ac.za
[2]nita.mennega@up.ac.za

* Mike Nkongolo

**Abstract.** This research paper adopts a methodology by conducting a thorough literature review to uncover the essential prerequisites for achieving a prosperous career in the field of Information Security (IS). The primary objective is to increase public awareness regarding the diverse opportunities available in the Information Security (IS) field. The initial search involved scouring four prominent academic databases using the specific keywords "cybersecurity" and "skills," resulting in the identification of a substantial corpus of 1 520 articles. After applying rigorous screening criteria, a refined set of 31 relevant papers was selected for further analysis. Thematic analysis was conducted on these studies to identify and delineate the crucial knowledge and skills that an IS professional should possess. The research findings emphasize the significant time investment required for individuals to acquire the necessary technical proficiency in the cybersecurity domain. Furthermore, the study recognizes the existence of gender-related obstacles for women pursuing cybersecurity careers due to the field's unique requirements. It suggests that females can potentially overcome these barriers by initially entering the profession at lower levels and subsequently advancing based on individual circumstances.

**Keywords:** Information security skills, cybersecurity competencies, digital security expertise, information security characteristics, cybersecurity behavioral attributes, digital security dispositions

## 1 Introduction

The demand for IS professionals has surged due to the escalating cyber-threats [1-5]. This study focuses on the shortage of skilled IS experts and the specific challenges faced by women's gender in entering this field. Additionally, it examines the diverse nature of cyber-threats, including their impact on data security, financial vulnerabilities, and instances of cyberbullying [6]. Recent research conducted by the IS Workforce has revealed a significant deficit of 3 000 000 IS professionals [7]. This scarcity is primarily attributed to the growing number and sophistication of cyber-threats. The proliferation of internet connectivity and the widespread use of wireless networks has expanded the attack surface for hackers, resulting in increased security breaches. The consequences of cyber-threats extend beyond financial impacts. Zero-day attacks, for example, can cause substantial data loss and financial harm. Furthermore, cyberbullying poses a psychological threat to individuals, as it can lead to embarrassment and potentially prompt drastic actions. Home automation, involving IoT devices, introduces another avenue for cyber-threats, allowing malicious actors to gain control over household systems. A concerning scenario includes unauthorized manipulation of residential gate controls by hackers [7]. While there is a shortage of IS professionals in general, women encounter distinct challenges in entering the field. Factors such as long and irregular working hours for maintenance windows contribute to the underrepresentation of women in the IS profession. However, there is growing recognition of the need to foster greater gender diversity in the IS landscape [8], as it brings unique perspectives and skills to address the evolving threatscape (Figure 1).

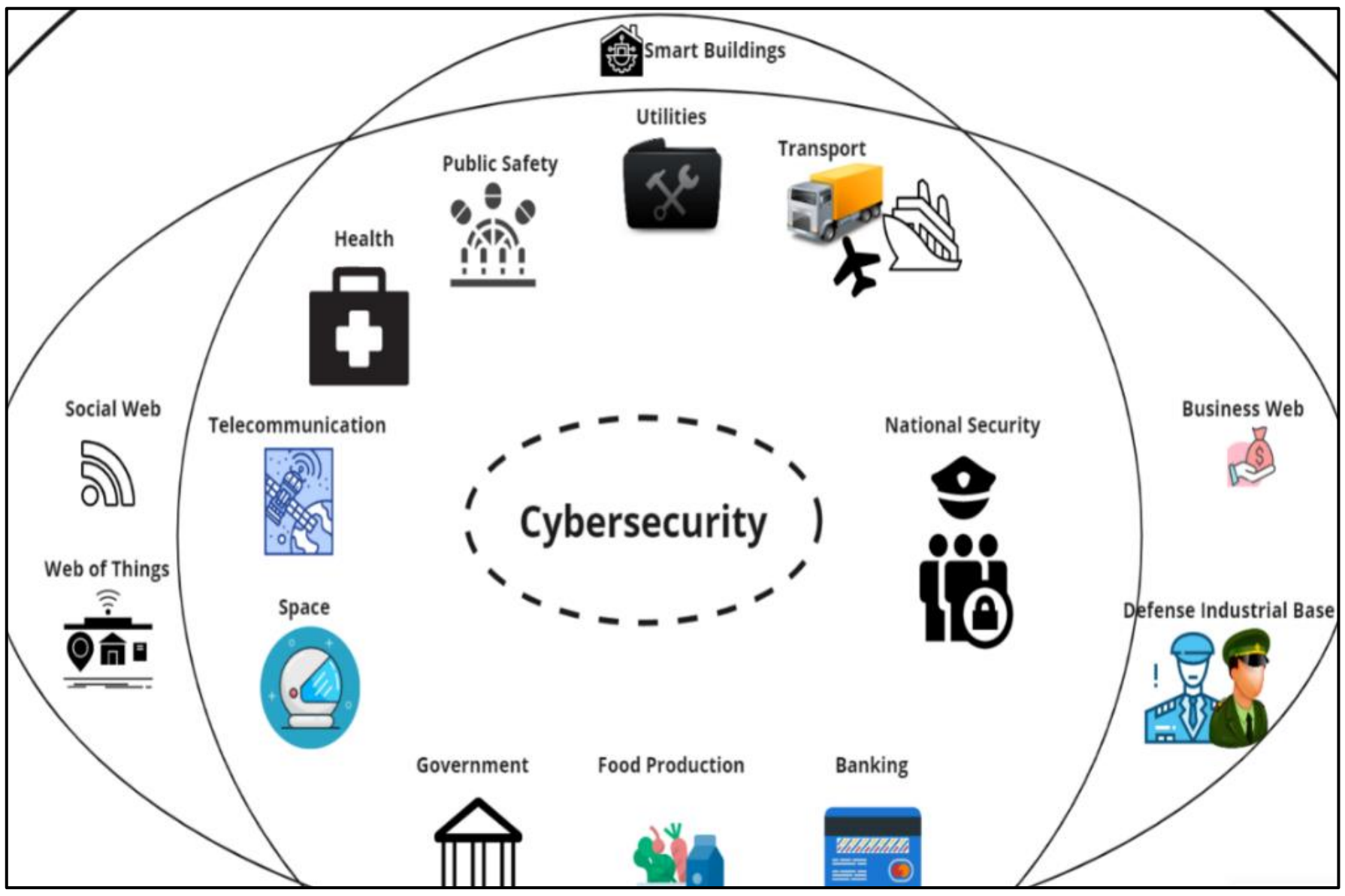

**Fig. 1.** The field of cybersecurity

To effectively establish and maintain robust cybersecurity systems, professionals in the field of Information Technology must possess a comprehensive understanding of the interplay between hardware and software components [9]. Experience and technological knowledge are crucial in developing the requisite skills necessary for success in the IS domain. Furthermore, individuals specializing in IS should embody specific qualities [10]. A comprehensive review of the literature was conducted to acquire a more profound comprehension of the requirements and potential challenges encountered by individuals aspiring to pursue careers in the field of IS. It is worth noting that the specific requirements may vary depending on the particular job specifications. Moreover, this study explores an approach to fostering IS awareness and encouraging greater participation in the IS field. The article is structured into distinct sections, with Section 2 providing a concise overview of the IS landscape. Section 3 delineates the research methodology employed in this study, while Section 4 presents the findings derived from the literature review. Finally, Section 5 concludes the study by succinctly summarizing the key findings and their implications.

## 2 Cybersecurity foundation

Numerous private and public organizations face significant security challenges arising from the utilization of wireless communications [11]. The repercussions of such actions can be severe, as they have the potential to disrupt governmental systems by introducing viruses that encrypt sensitive data. For instance, malicious codes can be deployed to instruct a National Database to crash, leading to disruptions in national security and causing substantial harm [11, 12]. A vishing attack tricks entities into divulging sensitive information [13]. According to reports from the Federal Bureau of Investigation (FBI), scams have emerged as some of the most prevalent crimes since 2021 [14]. In addition, the IC3 (Internet Crime Complaint Center) has reported an overwhelming number of complaints, exceeding 28 000, that is related to vishing themes. Given the seriousness of these challenges, the field of IS demands a diverse set of IS skills to effectively safeguard critical infrastructures against cyberattacks [16, 17].

### 2.1 Information security expertise

To safeguard critical infrastructures against an ever-growing array of threats, it is imperative to automate data privacy and security measures. Information security skills encompass a range of areas related to deep packet inspection [17, 18]. Additionally, soft skills such as critical thinking, adaptability, and communication play a crucial role in ensuring comprehensive IS measures. Within the realm of cybersecurity careers, limited research has been conducted on the IS landscape. However, a study by [19] identified several essential skills, including teamwork and ethical decision making. Another study [20] highlighted the significance of technical skills, specifically experience in detecting network attacks and proficiency in utilizing various products for deep packet inspection. Furthermore, [21] stated that many IS-related job positions emphasized the importance of prior work experience. In summary, a comprehensive approach to cybersecurity requires both technical expertise and a range of soft skills to effectively protect critical infrastructures from threats. The duties of an IS professional could encompass the following:

- Providing support at Tier 1, Tier 2, and Tier 3 levels for network Traffic Management Function (TMF) solutions.
- Conduct regular monthly reviews of client systems to identify and address any issues.
- Planning, coordinating, and implementing resolutions for the identified issues on client systems.

- Carrying out monthly backups of each client's configuration [23, 24].

Moreover, one needs to possess the capability to deploy diverse user-friendly interfaces aimed at monitoring network traffic and identifying potential malicious intrusions. The interface should be intuitive, while also offering advanced features that shows the necessary level of detail to monitor the network traffic [23, 24]. A comprehensive understanding of networking is crucial to detect potential network anomalies [24]. The company may undertake the responsibility of providing relevant training to the cybersecurity employee. This training could be conducted internally or, when necessary, through training programs offered by trusted vendor partners. However, it is worth noting that such vendor training programs often involve significant financial investments on the part of the company.

### 2.2 Duties and tasks of an IS professional

A key focus for the IS professional involves analyzing the network traffic to identify and thwart network threats. Their responsibilities encompass various tasks such as assisting in the implementation of deep packet inspection systems [25]. Data analysis and utilization of Machine Learning (ML) skills are essential for creating insightful reports and dashboards used for operational purposes to effectively identify and combat malicious intrusions [26]. Troubleshooting proficiency is crucial for identifying and resolving system issues and ensuring the smooth delivery of network services. Efforts should be made to enhance the efficiency, quality, and value of traffic management systems [27]. To succeed in this position, candidates must demonstrate a strong ability to think creatively and proactively in developing solutions that enhance intrusion detection.

### 2.3 Job nature and requirements

An IS professional's responsibilities include designing, developing, optimizing network systems, ensuring system reliability, and implementing new policies for network management. They utilize their knowledge to provide support, analyze statistics, and make recommendations for more efficient network attack detection [28, 29]. To excel in this role, one should have a strong understanding of networking protocols [29], proficiency in networked applications, familiarity with operating systems, skills in database administration, data analysis, and knowledge of artificial intelligence and machine learning.

## 3 Research Approach

To identify pertinent research articles, a systematic literature review methodology was employed. The research questions guiding this review are depicted in Figure 2.

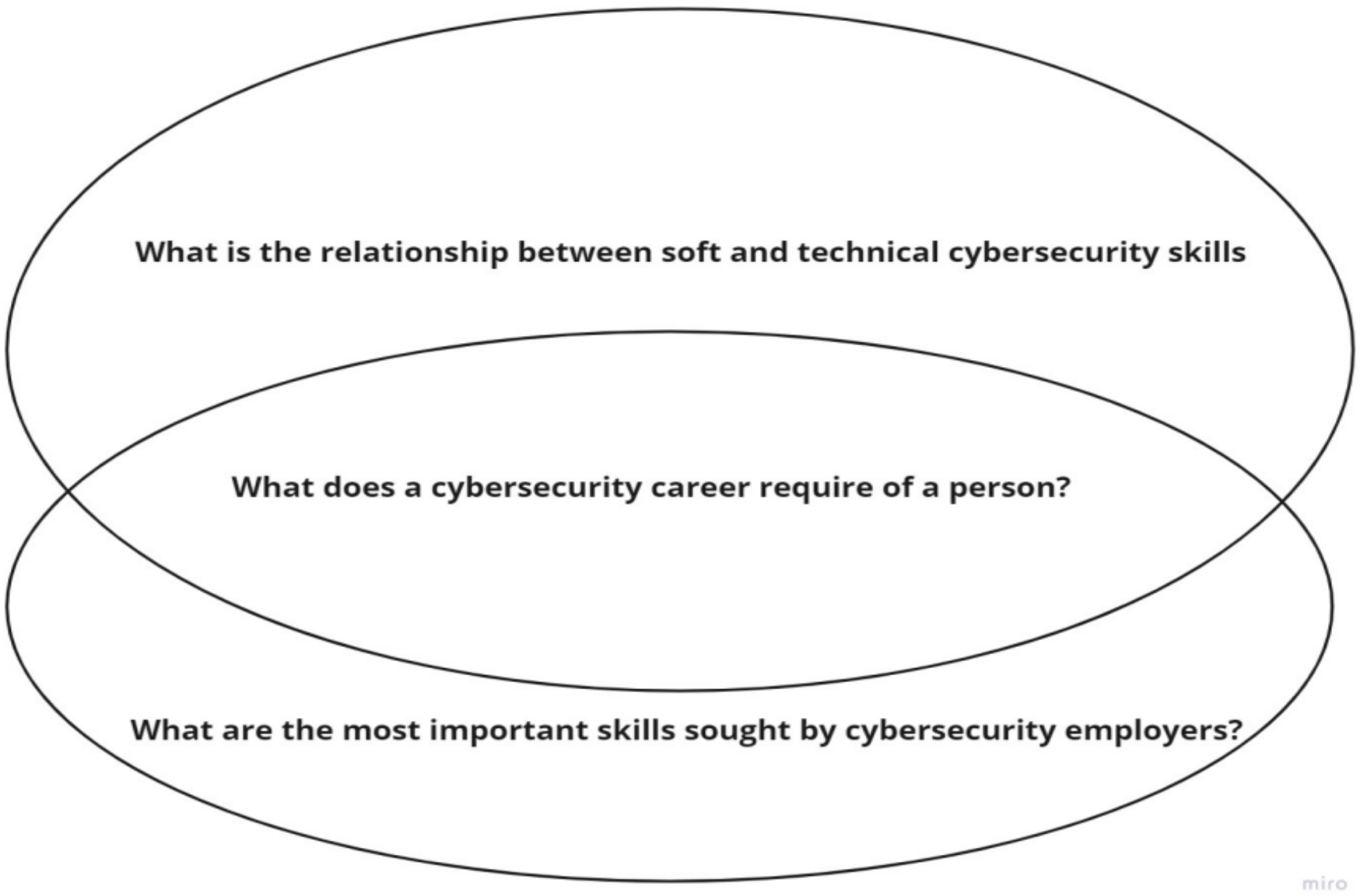

**Fig. 2.** The investigation queries

## 3.1 Words used to search

To find relevant published articles, the search was conducted using the keywords depicted in Figure 3.

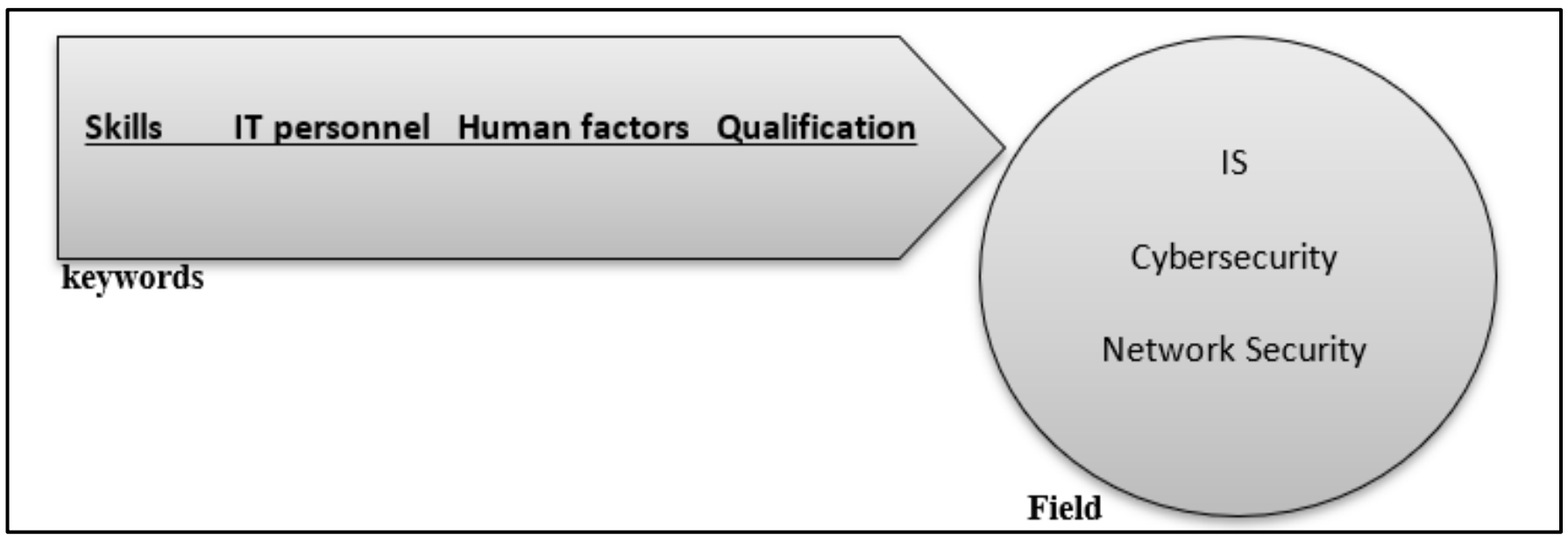

**Fig. 3.** The search terms employed to identify pertinent articles

## 3.2 Selection criteria

This study incorporated academic articles and conference proceedings that focused on IS topics.

### 3.1 Rejection criteria

Non-peer-reviewed and inadequately described articles were excluded from this study.

### 3.1 Data gathering

Figure 4 depicts the databases employed for data retrieval, while Figure 5 outlines the article selection process. Figure 5 displays the diagram employed in this research. The PRISMA diagram represents the "Preferred Reporting Items for Systematic Reviews and Meta-Analyses". This diagram enables authors to showcase the literature review's quality. Following the exploration of literature using the aforementioned keywords (depicted in Figure 3) and the acquisition of a total of 1 520 citations, we initially eliminated any duplicate entries. Subsequently, we evaluated the titles and abstracts of the articles to determine their relevance in addressing the research questions.

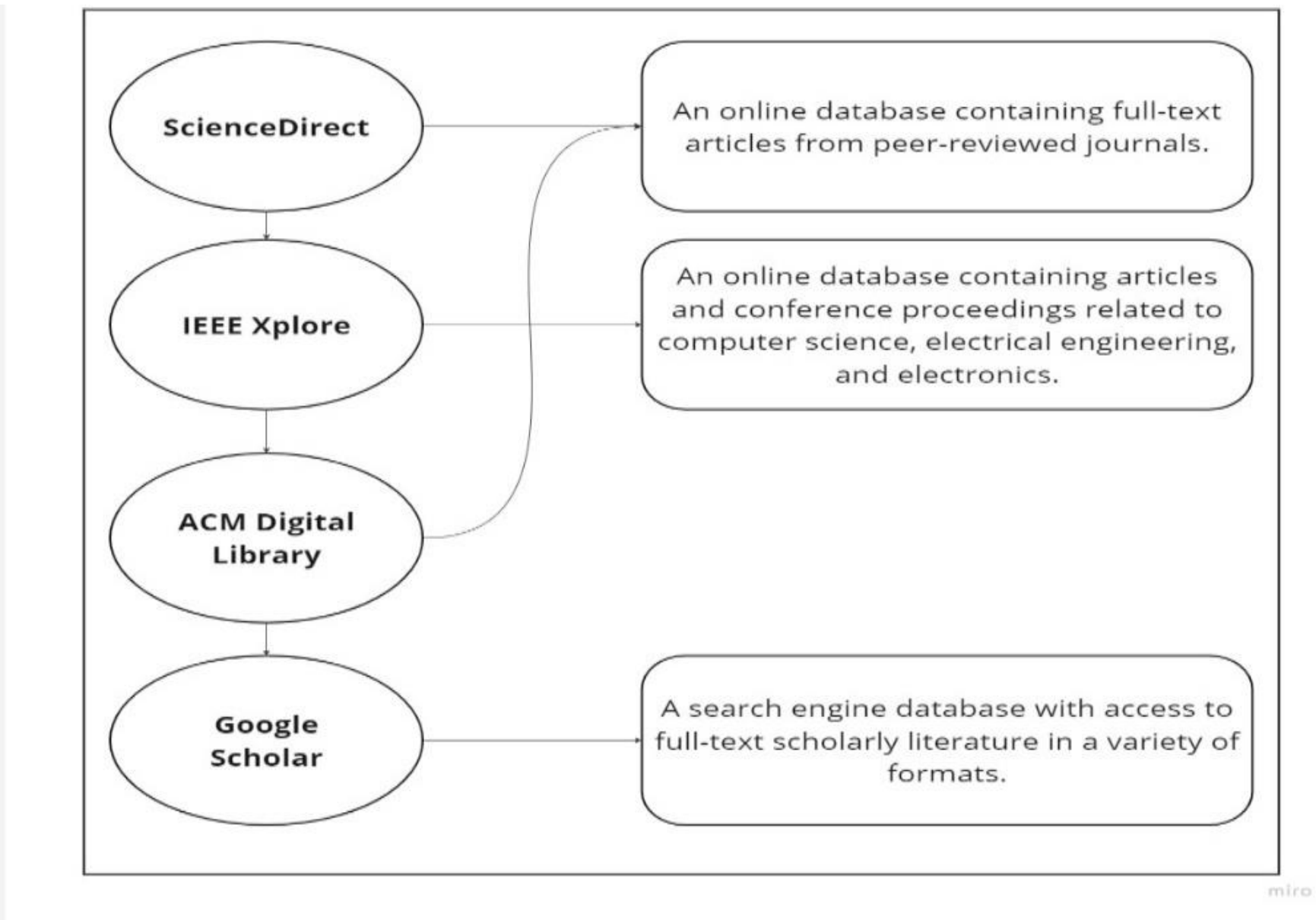

**Fig. 4.** Academic repositories used

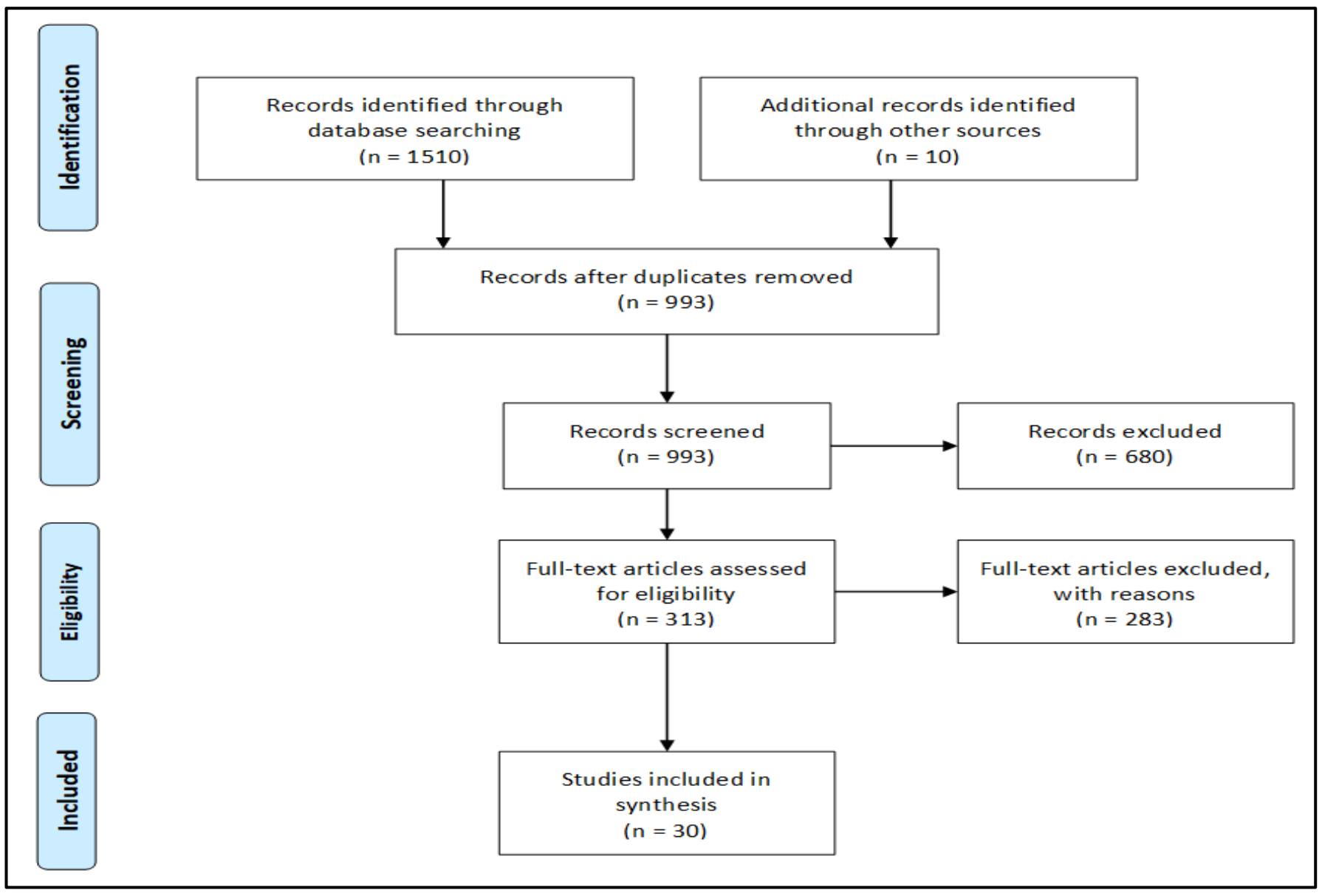

**Fig. 5.** The PRISMA flowchart depicting the article screening process

The search process identified 31 relevant research articles listed in Table 1. Figure 6 showcases a Wordcloud that highlights the key keywords extracted from the literature on cybersecurity. The analysis of Wordcloud indicates an increasing use of machine learning (ML) in the IS industry, highlighting the significance of ML skills for IS professionals. ML, a subset of artificial intelligence (AI), involves training algorithms to recognize patterns from data [30, 35]. In the field of IS, ML has diverse applications, particularly in threat detection and automated response. One notable application is malware classification [2, 3, 30], where ML classifiers assign scores to network traffic samples based on their maliciousness, representing the confidence level of the classification or prediction process.

### 3.5 Quality appraisal

The quality of the 31 research articles was assessed using four QA questions, and each article was given a subjective score to indicate its relevance. The QA questions are depicted in Figure 7, and the corresponding QA ratings for each paper are recorded in Table 2.

Table 2 demonstrates that four papers achieved a QA rating score of three or higher out of four, whereas three papers received a score of one or lower. These findings highlight the high relevance of the majority of the 31 final papers, suggesting their valuable contribution to address the research questions.

## 4 Results

Figure 8 showcases the publication trend of IS research starting from 2004. The figure reveals a consistent upward trajectory in the interest and focuses on cybersecurity over the years. Notably, there was a noticeable decrease in the number of publications during the year 2020, which can be attributed to the global shift towards remote work due to the COVID-19 pandemic.

This renewed surge indicates the continued importance and relevance of the subject in the field of research. Figure 9 presents a column chart depicting the distribution of published articles across various IS concepts.

The chart highlights that the most prominent concept explored in the articles was the skill set needed for an IS career [29]. Following closely behind were topics related to IS education and training, as well as the IS workforce.

The top contributors to published articles on IS are Australia and the United States. Addressing cyber-attacks that pose a threat to critical infrastructures remains a significant challenge. To effectively tackle these issues, IS professionals must possess specific skills and expertise [18].

Organizations can bridge the skills gap by implementing IS awareness training programs such as paper-based materials, interactive games, simulations, and video-based training programs.

**Table 1.** The finalized articles (Key: South Africa-SA, Australia-Aus)

| Nr | Citation | Country | Focus of paper |
|---|---|---|---|
| 1 | (Yair Levy, 2013) | USA | Skills needed for a career in cybersecurity. |
| 2 | (Von Solms & Van Niekerk, 2013) | SA | Threats and risks involved in the cyber environment and how it impacts individuals. |
| 3 | (Reeves et al., 2021) | Aus | Training is needed to acquire the appropriate skills for cybersecurity. |
| 4 | (Li et al., 2019) | USA | Training is needed to acquire the appropriate skills for cybersecurity. |
| 5 | (Hoffman et al., 2012) | USA | Training is needed to acquire the appropriate skills for cybersecurity. |
| 6 | (Hadlington, 2017) | UK | The focus is on the skills needed for a career in cybersecurity. |
| 7 | (Gratian et al., 2018) | USA | Intentions of an individual when working in a cyber domain. |
| 8 | (Furnell et al., 2017) | UK | Skills needed for a career in cybersecurity. |
| 9 | (Furnell, 2021) | UK | Skills needed for a career in cybersecurity. |
| 10 | (Dawson & Thomson, 2018) | USA | Workforce and the work environment of a cybersecurity specialist. |
| 11 | (Catota et al., 2019) | USA | Training needed to acquire the appropriate skills for cybersecurity. |
| 12 | (Caldwell, 2013) | USA | Skills needed for a career in cybersecurity. |
| 13 | (Burley et al., 2014) | USA | Workforce and the work environment of a cybersecurity specialist. |
| 14 | (Shahriar et al., 2016) | USA | Skills that need to be improved for a cybersecurity career. |
| 15 | (Jeong et al., 2019) | Aus | Personality of an individual working in a cybersecurity career |
| 16 | (Chowdhury et al., 2018) | Aus | Time pressure, how it impacts individuals with tasks to be completed in a certain timeframe. |
| 17 | (Paulsen et al., 2012) | USA | Workforce and the work environment of a cybersecurity specialist |
| 18 | (Bagchi-Sen et al., 2010) | USA | Skills needed for a career in cybersecurity |
| 19 | (Liu & Murphy, 2016) | USA | Education needed to acquire the appropriate skills for cybersecurity and which gender is mostly in the cybersecurity career. |
| 20 | (Javidi & Sheybani, 2018) | USA | Training needed to acquire the appropriate skills for cybersecurity. |
| 21 | (Sharevski et al., 2018) | USA | Training skills for cybersecurity. |
| 22 | (Besnard & Arief, 2004) | UK | Computer security and factors that influence cybersecurity. |
| 23 | (Martin & Rice, 2011) | Aus | Cybercrime and the impact it has on individuals and communities. |
| 24 | (Bauer et al., 2017) | USA | Making people and communities aware of the impact of cybercrime so that they are more careful with their information. |
| 25 | (Abawajy, 2014) | Aus | Make people aware of information security in different ways. |
| 26 | (Albrechtsen & Hovden, 2009) | Norway | Emphasize the different viewpoints between specialists and users of information security. |
| 27 | (Christopher et al., 2017) | USA | Identify the cybercrime trends and educate specialists. |
| 28 | (Baskerville et al., 2014) | USA, Italy | Identify and explain the threat paradigm versus the response paradigm of a cyber-attack. |
| 29 | (Rajan et al., 2021) | India, UK | Management of cybersecurity in an organisation. |
| 30 | (Hong & Furnell, 2021) | China, UK, SA | Behaviour of individuals exposed to cybercrime and how they respond to it. |
| 31 | (Kam et al., 2020) | USA | Learning required to improve skills of the cybersecurity workforce. |

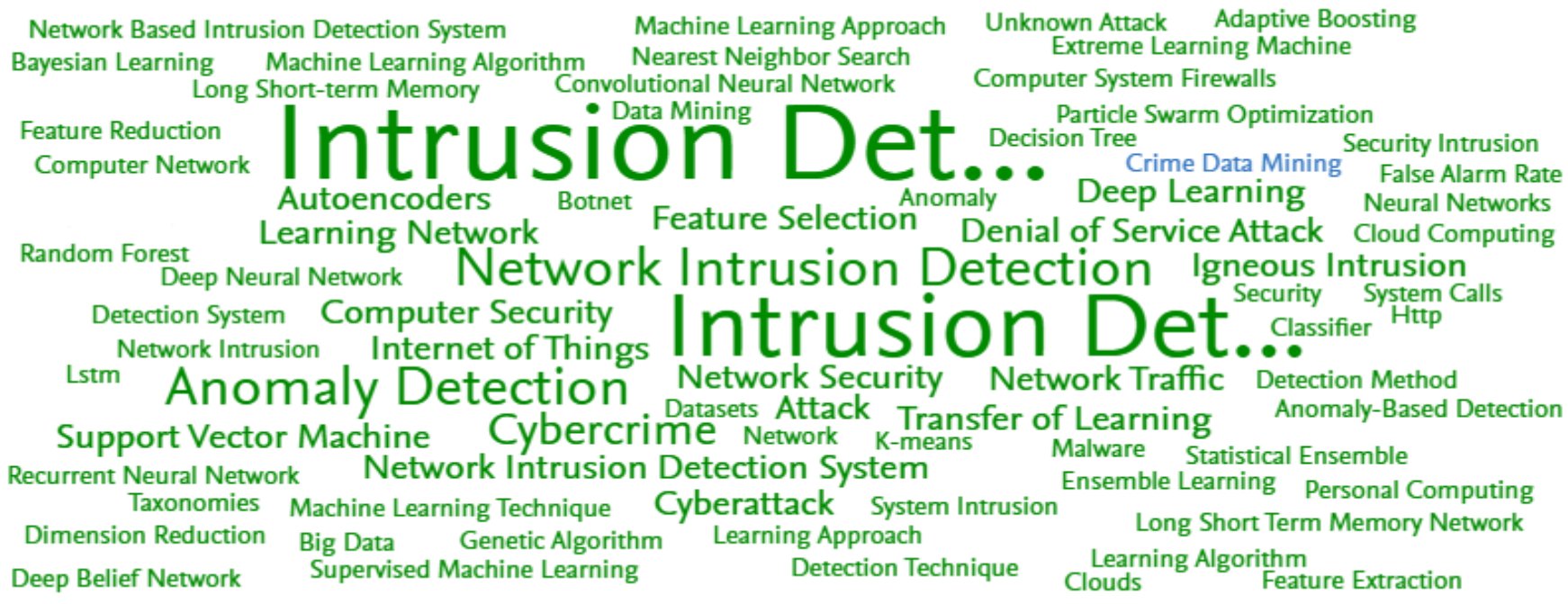

**Fig. 6.** The terms employed in the cybersecurity domain

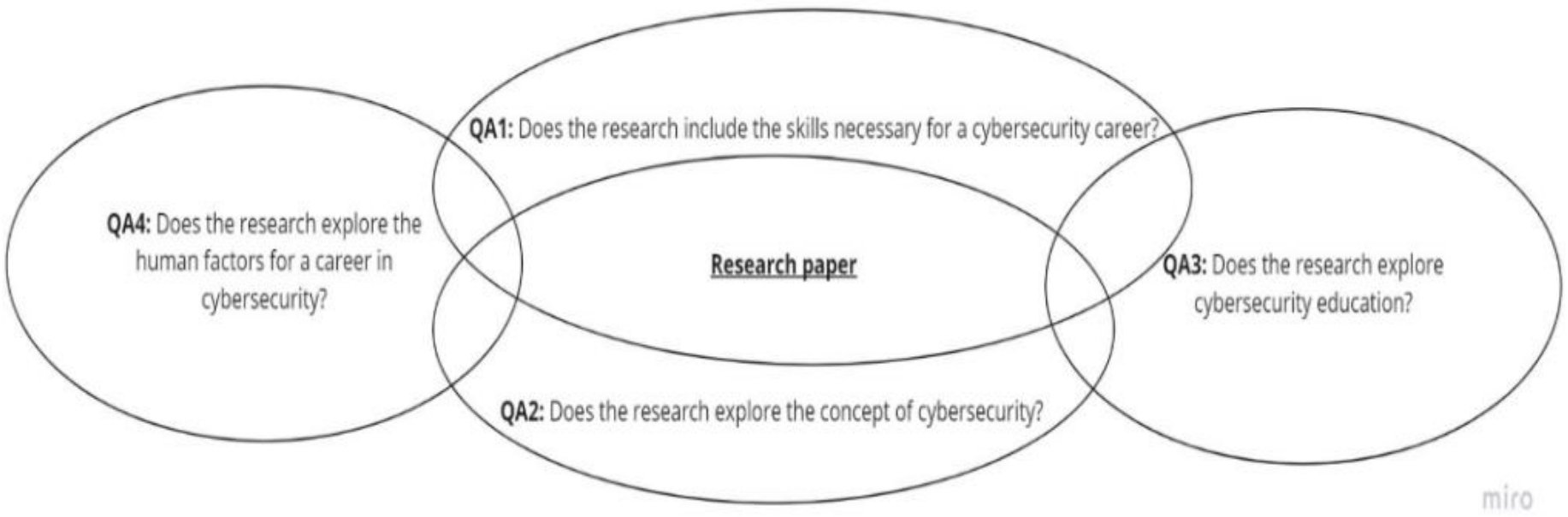

**Fig. 7.** The questions used to assess the quality of research articles

This study investigates the prerequisites for pursuing a career in IS and examines the relative importance of technical and soft skills.
The findings reveal that technical aptitude assumes a more significant role in this domain, with specific proficiencies like troubleshooting and familiarity with vendor products emerging as vital.
Moreover, the study highlights that full-time employment in IS typically encompasses an average of 40 working hours per week, occasionally necessitating after-hours assistance for clients.

## 5 Conclusion

This research brings novelty by conducting a comprehensive systematic literature review that examines the requirements, skills, and knowledge necessary for a career in IS. It employs rigorous methodologies, including specific keyword-based searches in bibliographic databases, to collect relevant data for analysis. Through thematic analysis, the study uncovers valuable insights into IS skills, education, training, and awareness initiatives, contributing to the understanding of the field. Notably, it sheds light on the demand for female professionals in IS and proposes policies and game theory strategies to enhance data security and privacy. The intention to develop an IS awareness game further demonstrates a novel approach to promoting data protection and privacy. Lastly, the emphasis on encouraging females to enter the field and progress through experience and qualifications highlights an important aspect of fostering diversity and inclusivity in the IS field.

**Table 2.** The ratings applied to the selected research papers

| Citation | QA1 | QA2 | QA3 | QA4 | Score |
|---|---|---|---|---|---|
| (Yair Levi, 2013) | Partial | Yes | No | Partial | 2.0 |
| (Von Solms & Van Niekerk, 2013) | No | Yes | No | Partial | 1.5 |
| (Reeves et al., 2021) | No | Yes | Yes | Partial | 2.5 |
| (Li et al., 2019) | No | Yes | No | Yes | 2.0 |
| (Hoffman et al., 2012) | Partial | Yes | Yes | No | 2.0 |
| (Hadlington, 2017) | No | Yes | No | Yes | 2.0 |
| (Gratian et al., 2018) | No | Yes | Partial | Yes | 2.0 |
| (Furnell et al., 2017) | No | No | Partial | No | 0.5 |
| (Furnell, 2021) | Yes | Yes | Partial | Partial | 3.0 |
| (Dawson & Thomson, 2018) | Yes | Yes | Partial | No | 2.5 |
| (Catota et al., 2019) | Yes | Yes | Yes | No | 3.0 |
| (Caldwell, 2013) | Partial | Yes | Partial | No | 2.0 |
| (Burley et al., 2014) | No | Yes | Partial | No | 1.5 |
| (Shahriar et al.) | No | Partial | No | No | 0.5 |
| (Jeong et al.) | No | Yes | Partial | Yes | 2.5 |
| (Chowdhury et al.) | No | Yes | Partial | Yes | 2.5 |
| (Paulsen et al., 2012) | Partial | Yes | Yes | No | 2.5 |
| (Bagchi-Sen et al., 2010) | Yes | Yes | Yes | Partial | 3.5 |
| (Liu & Murphy) | Yes | Yes | Partial | Partial | 3.0 |
| (Javidi & Sheybani) | Yes | Yes | Yes | Partial | 3.5 |
| (Sharevski et al.) | Partial | Yes | Yes | No | 2.5 |
| (Besnard & Arief, 2004) | No | Yes | Partial | No | 1.5 |
| (Martin & Rice, 2011) | No | Yes | No | No | 1.0 |
| (Bauer et al., 2017) | No | Yes | Partial | No | 1.5 |
| (Abawajy, 2014) | No | Yes | Partial | No | 1.5 |
| (Albrechtsen & Hovden, 2009) | Partial | Yes | No | No | 1.5 |
| (Christopher et al., 2017) | No | Yes | Partial | Partial | 2.0 |
| (Baskerville et al., 2014) | No | Yes | No | No | 1.0 |
| (Rajan et al., 2021) | No | Yes | Partial | No | 1.5 |
| (Hong & Furnell, 2021) | No | Yes | Partial | Yes | 2.5 |
| (Kam et al., 2020) | Partial | Yes | Yes | No | 2.5 |

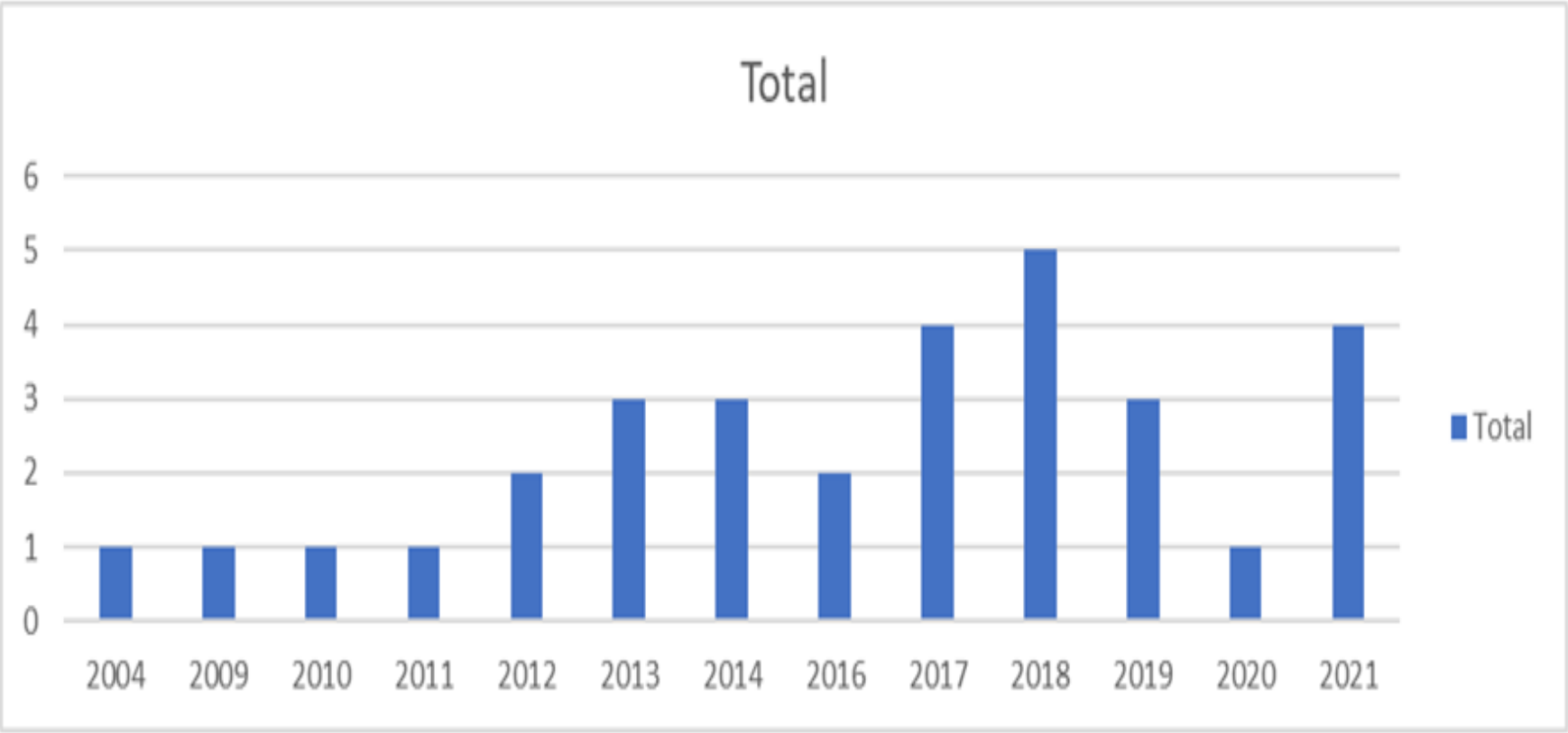

**Fig. 8.** Annual tally of cybersecurity publications

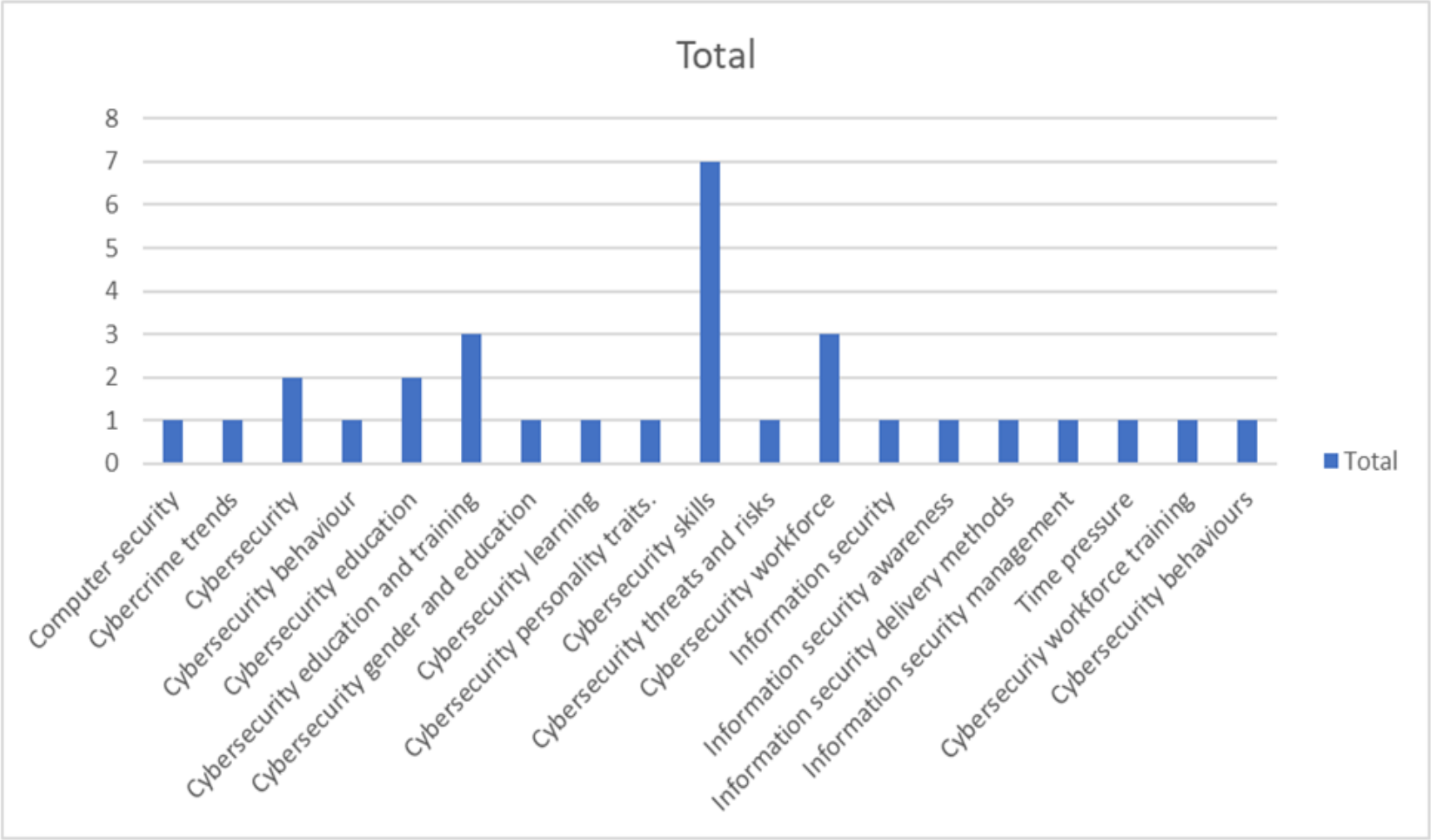

**Fig. 9.** Key themes extracted from the analyzed papers